%% file: main.tex
\pdfoutput=1

\documentclass[11pt]{article}

\usepackage{acl}

\usepackage{times}
\usepackage{latexsym}
\usepackage{enumitem}
\usepackage{url}

\usepackage[T1]{fontenc}

\usepackage[utf8]{inputenc}

\usepackage{microtype}

\usepackage{inconsolata}

\usepackage{xcolor}
\usepackage{tikz-dependency}
\usepackage{booktabs}
\usepackage[export]{adjustbox}
\usepackage{multirow}
\usepackage{bbding}
\usepackage{amsmath}

\usepackage{tabularx}

\usepackage{listings}
\usepackage{float}
\usepackage{subcaption}

\usepackage{soul}

\author{First Author \\
  Affiliation / Address line 1 \\
  Affiliation / Address line 2 \\
  Affiliation / Address line 3 \\
  \texttt{email@domain} \\\And
  Second Author \\
  Affiliation / Address line 1 \\
  Affiliation / Address line 2 \\
  Affiliation / Address line 3 \\
  \texttt{email@domain} \\}

\newcommand{\notesn}[1]{\textcolor{green}{\bf\small [#1 --SN]}}

\newcommand{\finqadataset}{FinQA}
\newcommand{\convfinqadataset}{ConvFinQA}
\newcommand{\modelname}{FINDER}

\title{Program of Thoughts for Financial Reasoning: Leveraging Dynamic In-Context Examples and Generative Retrieval}

\author{Subhendu Khatuya, Shashwat Naidu, Pawan Goyal, Niloy Ganguly \\
Indian Institute of Technology Kharagpur, India\\
    subha.cse143@gmail.com, shashwatnaidu07@gmail.com, \\  pawang@cse.iitkgp.ac.in, niloy@cse.iitkgp.ac.in
}

\begin{document}

\maketitle

\begin{abstract}

Despite continuous advancements in the capabilities of large language models (LLMs), numerical reasoning remains a challenging area.
Techniques like chain-of-thought prompting, tree-of-thought prompting, and program-of-thought prompting guide LLMs through intermediate reasoning steps. Although in-context learning with few-shot prompting has improved performance, LLMs still lag behind state-of-the-art models on financial numerical reasoning datasets such as FinQA and ConvFinQA. In this work, we introduce \modelname, a novel two-step framework, to enhance LLM's capabilities in financial numerical reasoning. The first step utilizes a generative retriever to extract relevant facts from unstructured data, including both text and tables. This is followed by context-aware Program of Thought prompting with dynamic selection of in-context examples. Our model \modelname~achieves a new state-of-the-art performance on both the \finqadataset~and \convfinqadataset~datasets, surpassing previous benchmarks with execution accuracy improvements of \textbf{5.98\% }and \textbf{4.05\%}, respectively.
\end{abstract}

\input{Sections/Introduction}

\input{Sections/RelatedWorks}
\input{Sections/Methodology}
\input{Sections/Dataset}
\input{Sections/Experimental_Setup}
\input{Sections/Results}

\input{Sections/Conclusion}
\input{Sections/Limitations}

\bibliography{main}

\appendix

   \input{Sections/Appendix}

\end{document}

%% file: Sections/Introduction.tex
\section{Introduction}

Numerical reasoning, particularly within financial domains, remains a significant challenge in artificial intelligence (AI). Unlike traditional question-answering tasks \cite{rajpurkar-etal-2018-know, yang-etal-2018-hotpotqa}, it requires not only the extraction of relevant information from diverse sources, such as tables and unstructured text, but also the construction of coherent reasoning paths to integrate and process this information. Towards this effort, datasets such as FinQA \cite{chen-etal-2021-finqa} and ConvFinQA \cite{chen-etal-2022-convfinqa} have been developed to benchmark deep learning models for such numerical reasoning tasks in the financial domain.

Despite advances in task-specific models, large language models (LLMs) still struggle with numerical reasoning \cite{huang2022towards, satpute2024can, zhao-etal-2024-docmath}, as it requires multi-step problem-solving including fact extraction, logical inference, and mathematical computation. Even minor errors in intermediate steps can lead to incorrect solutions, making numerical tasks particularly challenging for LLMs.

A common approach to tackle numerical reasoning problem is the retriever-generator question-answering framework, initially proposed by \citet{chen-etal-2021-finqa}. Subsequent models leveraging pre-trained language models have shown significant performance gains on such tasks \cite{wang2022novel, zhang2022robustly, wang2022numerical}. 
The current state-of-the-art is the APOLLO \cite{sun2024apollo} model, a retriever-generator framework that employs a number-aware retriever and a fine-tuned BERT-based encoder-decoder generator to generate executable programs, achieving high execution accuracy.

In parallel, an emerging line of work adopts the Program-of-Thoughts (PoT) paradigm \cite{chen2023program} for program generation, which utilizes decoder-only large language models to express the reasoning process as executable programs, disentangling computation from reasoning.These methods typically rely on in-context learning (ICL) rather than fine-tuning, offering greater flexibility and improved generalization capabilities. PoT-based approaches are more adaptable and better leverage the capabilities of modern LLMs.

In this work, we bridge these two directions by proposing a novel retriever-generator framework \modelname~ that replaces the retriever with an instruction-tuned generative model and the encoder-decoder based program generator with a decoder-only LLM under the PoT paradigm with some key modifications. PoT prompting \cite{chen2023program}, when applied to financial numerical reasoning tasks, typically includes the entire textual and tabular context in the prompt. This often leads to grounding errors, as the model struggles to accurately identify and extract the relevant numerical information. To address this, we instruction-tune FLAN-T5 \cite{chung2022scaling} using LoRA \cite{hu2021lora} to enable accurate instance-specific relevant facts extraction.


Additionally, PoT relies on static in-context examples, which may not generalize well across diverse problem instances. To address this limitation, we make use of dynamic selection of in-context examples. We enhance an existing gradient-based method, PromptPG \cite{lu2023dynamic} by refining both the candidate selection strategy and the reward evaluation mechanism which reduces the chance of incorrect penalties. We ensure candidate diversity \cite{rubin-etal-2022-learning}, by identifying a fixed-size subset of training samples from a larger pool. This subset is carefully constructed using clustering techniques to include representative questions from a range of themes (e.g., growth rates, amortization). 
In addition to this, we place higher-ranking examples closer to the query to optimize contextual relevance. Finally, we combine the dynamically selected diverse ordered in-context examples with the retrieved facts and format them using PoT-based prompting to get the final answer of the question.

\modelname\footnote{Data and code are available at \url{https://github.com/subhendukhatuya/FINDER_POT_Financial_Numeric_Reasoning.git}} surpasses existing LLM-based approaches with the PoT paradigm, achieving a \textbf{8.56\%} improvement on FinQA and a \textbf{9.60\%} improvement on ConvFinQA, compared to the previous best. It sets a new SOTA with execution accuracies of \textbf{75.32\%} on FinQA and \textbf{81.95\%} on ConvFinQA, improving upon the current SOTA APOLLO \cite{sun2024apollo} by \textbf{5.98\%} and \textbf{4.05\%}, respectively.

\if{0}
\notesn{

A common approach to tackle this problem is the retriever-generator question-answering framework, initially proposed by \citet{chen-etal-2021-finqa}. Subsequent models leveraging pre-trained language models have shown significant performance gains on such tasks \cite{wang2022novel, zhang2022robustly, wang2022numerical}. The current state-of-the-art is the APOLLO \cite{sun2024apollo} model, a retriever-generator framework that employs a number-aware retriever and a fine-tuned BERT-based encoder-decoder generator to generate executable programs, achieving high execution accuracy.

Despite significant progress in retriever-generator based frameworks, large language models (LLMs) based solutions continued to face challenges in numerical reasoning \cite{huang2022towards, satpute2024can, zhao-etal-2024-docmath}, as it demanded multi-step problem-solving involving fact extraction, logical inference, and precise mathematical computation. Even minor errors in intermediate steps could lead to incorrect solutions, making numerical tasks particularly challenging for LLMs. To mitigate these limitations, recent line of work have adopted the Program-of-Thoughts (PoT) paradigm \cite{chen2023program}, which expresses reasoning as executable programs using decoder-only LLMs, thereby disentangling reasoning from computation. These approaches typically rely on in-context learning (ICL) rather than fine-tuning, offering improved flexibility and generalization across tasks. However, despite their promise, such prompting-based methods often fall short of the execution accuracy achieved by the retriever-generator based works.

One major source of errors in financial numerical reasoning using LLMs is grounding errors, where the LLM struggles to identify and extract only the information necessary for solving the problem. This often stems from prompting strategies that include the entire textual and tabular context, overwhelming the LLM. This motivated us to bridge the two directions of solutions by proposing a novel retriever-generator framework \modelname~ that replaces the retriever with an instruction-tuned generative model and the encoder-decoder based program generator with a decoder-only LLM under the PoT paradigm with some key modifications. We instruction-tune FLAN-T5 \cite{chung2022scaling} using LoRA \cite{hu2021lora} to enable accurate instance-specific relevant facts extraction. 
}
\fi

%% file: Sections/RelatedWorks.tex
\section{Related Works}
Numerical reasoning in math word problems has long been a challenging task, addressed in several works \cite{huang2024joint, sun2024enhancing, liu-etal-2020-reasoning}. Several benchmark datasets have been developed to advance research in this area, including MathQA \cite{amini-etal-2019-mathqa} and MaWPS \cite{koncel-kedziorski-etal-2016-mawps}. Additionally, \newcite{lu2023dynamic} introduced the Tabular Math Word Problems (TABMWP) dataset, which requires mathematical reasoning over both textual and tabular data. Math word problems (MWP) are often presented in a structured and consistent format, prompting some prior research to use template-based approaches \cite{10.1609/aaai.v33i01.33017144} or tree-based methods \cite{jie-etal-2022-learning,li-etal-2023-trea} to tackle these problems more effectively. Beyond MWPs, long-form numerical reasoning is addressed by several state-of-the-art methods.

FinQA \cite{chen-etal-2021-finqa} and ConvFinQA \cite{chen-etal-2022-convfinqa} are benchmark datasets for long-form numerical reasoning over financial reports. Various approaches have been proposed, including fine-tuning and prompting-based methods. \newcite{wang2022numericalreasoningquestionanswering} pre-trained DeBERTa \cite{he2023debertav3improvingdebertausing} on financial data. Ant Risk AI \cite{zhang2022robustly} used an ensemble of specialized models, while ELASTIC \cite{wang2022numerical} introduced a cell retriever for extracting relevant gold cells. TabT5 \cite{andrejczuk-etal-2022-table} leveraged a T5 model pretrained on Wikipedia tables for reasoning tasks.  LLMs enhance long-form reasoning, with Chain of Thought (CoT) \cite{wei2022chain} and Program of Thought (PoT) \cite{chen2023program} prompting improving reasoning and prompt efficiency. \cite{lim-etal-2024-enhancing} enhanced argument recognition in program generation using an argument aggregator. CBR-Ren \cite{feng2024cbr} combines LLMs with case-based reasoning to enhance retriever-generator models by improving critical fact retrieval. 


%% file: Sections/Methodology.tex
\section{Methodology}
\label{methodology}
\begin{figure*}
    \centering
    \includegraphics[width=1\linewidth]{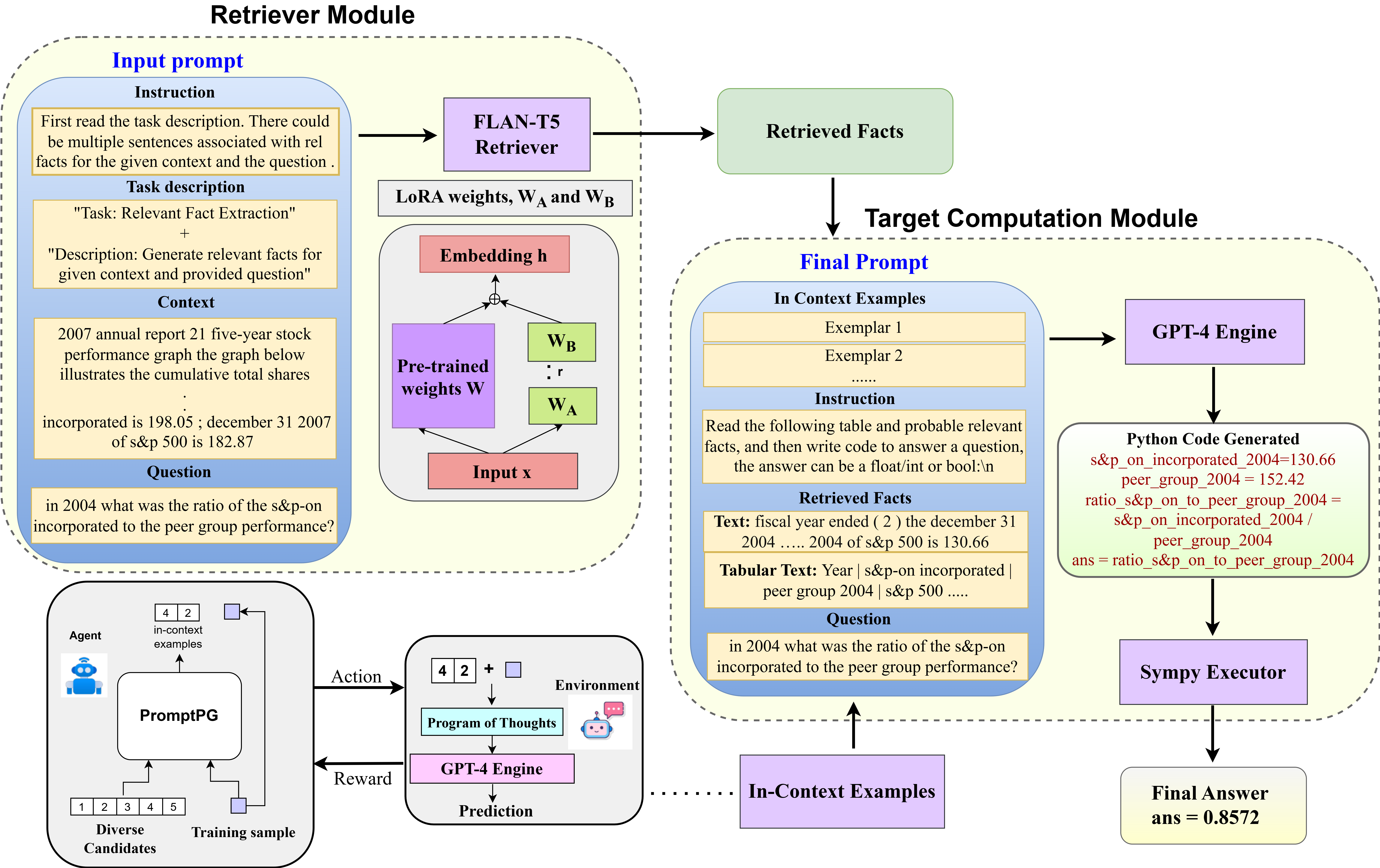}
    \caption{Architecture diagram of our proposed framework \modelname. FLAN-T5 processes task-specific instructions, context (text and tables), and questions to generate relevant facts. GPT-4 generates PoT style code using a Final Prompt with in-context examples from trained PromptPG and retrieved relevant facts. The SymPy executor then interprets this code to produce the final answer. }

    \label{fig:model diagram}
\end{figure*}

Our proposed framework for financial numerical reasoning task is divided into two phases, a generative relevant fact retriever phase, and a target answer computation phase as depicted in Figure \ref{fig:model diagram}.     

\subsection{Task Formulation}

In this task, the input consists of a structured table $T$ containing rows and columns of data, textual information, and a natural language question $Q$. The objective is to generate an output $Y$ that correctly answers the question by leveraging the information encoded within the table and the textual content. Given the complexity of this task, we break it down into two primary components.

\noindent \textbf{Relevant Fact Retriever:} This involves identifying the relevant facts from the input data.

\noindent \textbf{Target Answer Computation:} This involves developing and executing a program using the retrieved facts, in-context examples and the given question to derive the final answer.

\subsection{Relevant Fact Retriever} 

We use fine-tuning based retrieval methods to extract the relevant facts for the question. 


\subsubsection{Motivation} 
The current state-of-the-art method, APOLLO, relies on scoring-based retrieval with fixed line selection and task-specific number sampling strategies, which limit its flexibility and generalizability to broader settings. This motivated us to utilize finetuning based retriever as it offers inherent generalizability and can be seamlessly adapted to different LLM-based models.

 \subsubsection{Generative Retrieval}
 \label{subsubsec:gen_retrieval}
We fine-tune the FLAN-T5 Large model~\cite{chung2022scaling} to effectively retrieve relevant facts. Our selection of this large language model is based on its extensive pretraining through instruction tuning across a wide variety of tasks. FLAN-T5 Large, which has been instruction-tuned on over 1,800 tasks, demonstrates superior zero-shot and few-shot performance compared to its non-instruction-tuned counterpart, T5 Large \cite{raffel2023exploring}. Additionally its smaller scale compared to other LLM's makes it a more efficient choice.

The input prompt format is illustrated on the left side of Figure \ref{fig:model diagram}, consisting of an Instruction, Task Description, Context, and Question. We conduct the fine-tuning in a parameter-efficient manner using Low-Rank Adaptation (LoRA) \cite{hu2021lora}. Formally, the model is fine-tuned with the instruction prompt \( \mathbf{IP} \), containing a natural language description of the task, the table in textual format \( \mathbf{TabText}^i \), the textual data \( \mathbf{D}^i \), and the question $Q^i$. The modified input \( S_i \) for the \( i^\text{th} \) sample is represented as \( S_i = \mathbf{IP} \, || \, \mathbf{TabText}^i \, || \, \mathbf{D}^i \, || \, \mathbf{Q}^i \).
where \( || \) denotes text concatenation. During training, the model is optimized to produce the ground truth relevant fact. In the inference phase, it generates the retrieved facts \( \mathbf{RetFact} \) for the test samples. 
To enhance fact completeness, we apply the following  post-processing steps which also ensure factual consistency and mitigate hallucinations.

(1). We utilize Sentence-BERT \cite{sentencebert} to generate embeddings for each sentence in \textbf{RetFact} and in the context.
(2). For each sentence in \( \mathbf{RetFact} \), we select the most similar context sentence based on cosine similarity. The matched set is denoted as \( \mathbf{RetFactMatched} = \{r_1, r_2, r_3, \ldots\} \) where each \( r_i \) corresponds to the context sentence with the highest similarity to the \( i^\text{th} \) sentence in \( \mathbf{RetFact} \).

Unlike scoring-based methods that rely on fixed
thresholds for fact selection, our generative retrieval generates relevant facts dynamically, enabling more precise and context-aware extraction.


   

\paragraph{Comparative Experiments:}  To evaluate the effectiveness of our approach, we also experiment with the generative retriever Mistral-7B~\cite{jiang2023mistral} and the score-based retrieval module from APOLLO~\cite{sun2024apollo}. The results of these experiments are presented in Section~\ref{subsec:retriever_performance}.  

\subsection{Target Answer Computation}
The initial step in this stage involves the dynamic selection of in-context examples. We use Program of Thought (PoT) style prompting to generate Python code by GPT-4 \cite{openai2024gpt4}, which is then executed by an external interpreter to obtain the final results. Unlike the static prompts used in PoT, we leverage dynamic prompting. 

\subsubsection{Dynamic In-Context Example Selection}
\label{sssec:dynamic_in_context}
  We adapt the PromptPG framework \cite{lu2023dynamic} for our datasets to dynamically select in-context examples, introducing key modifications to its candidate selection strategy and reward evaluation mechanism. 

\noindent \textbf{Clustering of Questions:}
We first embed sentences using Sentence BERT~\cite{sentencebert} and then apply agglomerative clustering~\cite{mullner2011modern} to question embeddings, forming 50 clusters. For each cluster, we select a representative question by choosing the one that is nearest to the centroid. While we also tested with TF-IDF vectors, Sentence BERT yielded the best clusters with higher silhouette score \cite{shahapure2020cluster}. 

We used agglomerative clustering with average linkage and determined the optimal number of clusters by varying it between 30 and 70. Based on Silhouette Scores and qualitative assessment, 50 clusters provided the best balance between semantic coherence and conceptual diversity. This setting also aligned well with the coverage of distinct financial concepts in the dataset.

\begin{figure}[ht]
        \includegraphics[scale=0.20]{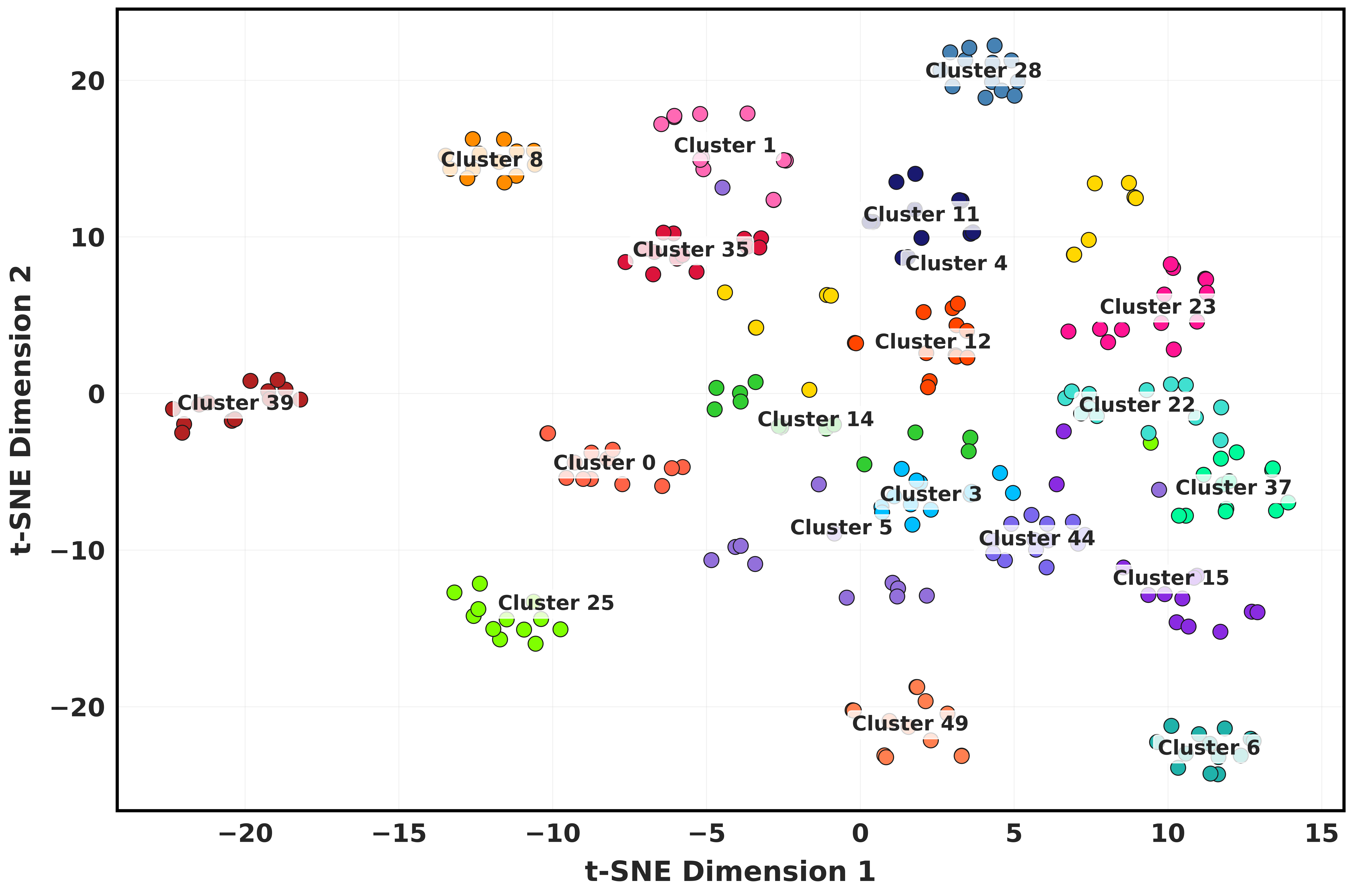}
        \caption{t-SNE plot illustrates the clusters formed from the embeddings of the training questions, showcasing a subset of clusters for visualization purposes. Each cluster represents a theme e.g. growth rates, amortization, percentage change, ROI, etc. 
} 
    
        \label{fig:question_clustering}
\end{figure}

Figure~\ref{fig:question_clustering} shows t-SNE plot for visualization of some clusters. In Table ~\ref{tab:clustered_questions}, we illustrate the clustering of semantically similar questions (showcased only two questions) with corresponding representative questions. Each cluster captures a specific theme, such as growth rates, percentage changes or ROI, with the representative question capturing the core intent of the group. Clustering helps the model identify patterns, generalize across contexts, and provide diverse in-context examples while maintaining a balanced candidate subset.

\begin{figure*} 
    \centering
     \includegraphics[width=\linewidth]
    {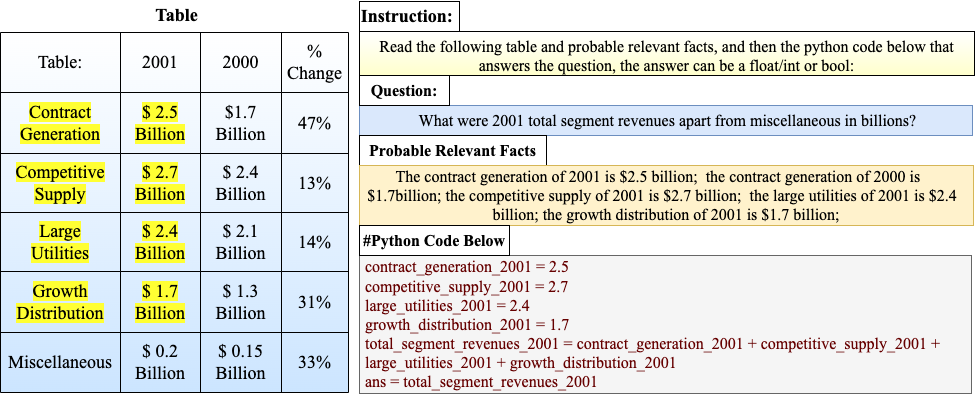}
     \caption{In-Context example formatted in Program of Thought Prompt style for FinQA.}
    \label{fig:few_shot_example}
\end{figure*}

Formally, given a problem $p_i$, we want the agent to find $K$ in-context examples $e_i = \{e_i^1, e_i^2, ..., e_i^K\}$ using policy gradient techniques from a candidate pool $E_{\text{cand}}$, and generate the answer $\hat{a}_i$, maximizing a reward $r_i = R(\hat{a}_i|p_i)$ with the help of GPT-4 engine. The in-context examples are selected according to a policy
\begin{equation}
\begin{aligned}
e_i^k &\sim \pi_{\theta}(e_i|p_i), \quad e_i^k \in E_{\text{cand}}, \\
    e_i^k &\text{ are independent for } k = \{1, 2, ..., K\}
\end{aligned}
\end{equation}
where $\theta$ are the policy’s parameters. First, Python code is generated as $c_i = \text{GPT-4}(e_i, p_i)$ using a PoT-style prompt. The final answer is computed as $\hat{a}_i = \text{exec}(c_i)$, where exec denotes execution of the Python code. The reward is then computed by evaluating the generated answer $\hat{a}_i$ with respect to the ground truth answer $a_i$:
\[
r_i = R(\hat{a}_i|p_i) = \text{SCORE}(\hat{a}_i, a_i), \quad r_i \in \{-1, 1\}
\]
The function $\text{SCORE}()$ returns 1 if the generated answer aligned with the label and -1 otherwise. During training, the goal is to maximize the expected reward of the generated answer under the policy: ${E}_{e_i \sim \pi_{\theta}(e_i|p_i)}[R(\text{exec}(\text{GPT-4}(e_i, p_i)))]$. The reward function is optimized using the Policy Gradient method \cite{Sutton1998}. Since the expected reward cannot be computed analytically in closed form, we estimate it using Monte Carlo sampling. To update the policy, we apply the REINFORCE algorithm \cite{reinforce}.
\paragraph{Enhancements Over Existing framework:}  
We summarize our modifications over existing work:  

\noindent\textbf{Diverse Candidate Questions:} We cluster the training questions and select a representative from each cluster to ensure diversity while keeping the subset of candidates concise.\\

\noindent\textbf{Reward Computation:} Unlike the existing method that generates and matches answers directly from GPT-4, we use PoT style prompting. The generated code is executed to obtain the final answer, evaluating examples based on their reasoning contribution, not penalizing for potential mathematical errors by GPT-4.

\subsubsection{Final Answer Computation}
We utilize Program of Thought (PoT) prompting \cite{chen2023program} to generate code for answering questions. In PoT, language models generate reasoning steps as Python programs, allowing the model to focus exclusively on the reasoning process, which enhances performance. Building on this, we employ our dynamic in-context example selection policy as elaborated in Section \ref{sssec:dynamic_in_context} to retrieve the most relevant examples for each question. These selected examples are formatted as shown in Figure \ref{fig:few_shot_example} and incorporated into the prompt. As shown in "\textbf{Final Prompt}" in Figure \ref{fig:model diagram}, the question, relevant facts retrieved by the retriever and the selected in-context examples are provided to the GPT-4 engine to generate python code. SymPy-based code executor \cite{sympy} is used to decode the code and achieve final answer. For ConvFinQA, we include prior questions in the prompt to provide conversational context before the target question for reasoning.


\begin{table}[ht]

\centering
\scriptsize
\resizebox{\linewidth}{!}{

\begin{tabular}{p{0.45\linewidth} | p{0.25\linewidth}}
\toprule
\textbf{Clustered Questions} & \textbf{Representative Question} \\
\midrule
    What is the growth rate in consolidated revenues from 2016 to 2017? 

    What is the growth rate of net sales from 2014 to 2015? 
    
     What is the growth rate in net revenue in 2015 for Entergy Louisiana? 

& What is the \textbf{growth rate} in the average price of the purchased shares from october to november 2014 \\
\hline
    What was the percentage change in the redeemable non-controlling in 2012? 
    For 2017, what was net interest income on average managed interest-earning assets in US\$? 
    
    What is the percentage change in discounted liabilities from 2013 to 2014? 

& What is the \textbf{percentage change} in the balance of level 3 investments assets from 2007 to 2008? \\
\bottomrule
\end{tabular}
}

\caption{Clusters of potential candidate examples and their Representative Examples} 

\label{tab:clustered_questions}
\end{table}

%% file: Sections/Dataset.tex
\section{Dataset \& Evaluation Metrics }
\label{dataset description}
We perform our experiments on the following two datasets:
\textbf{FinQA} \cite{chen-etal-2021-finqa} is a numerical reasoning dataset over long-form financial data, comprising 8,281 expert-annotated QA pairs from financial reports. It follows a 75\%/10\%/15\% split into 6,251 training, 883 development, and 1,147 test instances.

\textbf{ConvFinQA} \cite{chen-etal-2022-convfinqa} is a dataset of conversational long-form numerical reasoning over financial data with 2,715 simple and 1,177 hybrid conversations, divided into 3,037/421/434 for train/dev/test sets. For this dataset, we report results on the development set, as the test set requires program submissions in a specific format, incompatible with our python based outputs.

\noindent \textbf{Evaluation Metrics:}
We use execution accuracy as our evaluation metric following prior work.
 
\section{Baselines}

We compare the performance of our proposed model against several competitive models. 

\noindent \textbf{Current State-of-the-Art: }
\textbf{APOLLO} \cite{sun2024apollo}: Uses number-aware sampling for retrieval and encoder-decoder based model for generation.
    
\noindent \textbf{Prompting methods:} (1) \textbf{BloombergGPT}~\cite{wu2023bloomberggpt}: LLM designed for financial tasks. (2) \textbf{Codex} \cite{codex} (3) \textbf{GPT-3} \cite{brown2020language} (4) \textbf{GPT-3.5-turbo} \cite{NEURIPS2022_b1efde53} all evaluated in zero-shot setting (5) \textbf{CBR-Ren} \cite{CBR-Ren}.

\noindent \textbf{Program of Thought (PoT)} \cite{chen2023program}: Reported results are from PoT-Codex and PoT-SC-Codex. Running PoT with GPT-4 yielded 69.38\% accuracy, below the reported 74\% (Appendix \ref{apn:pot}).

\noindent\textbf{Chain of Thought (CoT)}~\cite{wei2022chain} We report Direct (in zero shot setting) and CoT (using chain of thought strategy) for GPT and codex variants taken from \cite{chen2023program}.

\noindent \textbf{Fine-Tuning-Based methods:} We compare the performance with below mentioned models.

    \noindent (1) \textbf{FinQANet} \cite{chen-etal-2021-finqa}, (2) \textbf{NeRd} \cite{ran-etal-2019-numnet}, (4) \textbf{Longformer}  \cite{beltagy2020longformer},  (4) \textbf{GPT-2} \cite{radford2019language}, (5) \textbf{T5} \cite{raffel2020exploring},  (6) \textbf{CellRetriever+UniLM} \cite{NEURIPS2022_522ef98b}, (7) \textbf{ELASTIC} \cite{wang2022numerical}, (8) \textbf{TabT5} \cite{andrejczuk-etal-2022-table}, (9) \textbf{DyRRen} \cite{jie-etal-2022-learning}, (10) \textbf{ENCORE} \cite{wang-etal-2024-enhancing-numerical} (11) \textbf{ArgRecog} \cite{lim-etal-2024-enhancing}.
    
\noindent \textbf{Human Performance:} Includes performance of both experts and non-experts on FinQA taken from \cite{chen-etal-2021-finqa}.

%% file: Sections/Experimental_Setup.tex
\section{Experimental Setup}
We conduct all experiments on a Tesla V100 32GB GPU, using pre-trained checkpoints from the \textit{Huggingface} Library\footnote{\url{https://huggingface.co/}} for FLAN-T5-Large (780M parameters) and Mistral-7B Instruct v0.2 (7B parameters). FLAN-T5-Large is instruction-tuned using the LoRA paradigm (effective parameters $\approx$0.59M) for 10 epochs (15 minutes/epoch) with a batch size of 8, learning rate ($\text{lr}$) $4e{-4}$, and rank 2. Mistral-7B is tuned in 4-bit precision with LoRA (effective parameters $\approx$3.4M) for 5 epochs (9 hours/epoch) using a batch size of 4, $\text{lr}$ $5e{-4}$, and rank 2.

%% file: Sections/Results.tex
\if{false}
\begin{table}[!th]

     \scalebox{0.85}{
       \begin{tabular}{p{3.7cm}|>{\centering\arraybackslash}p{1.7cm}|>{\centering\arraybackslash}p{1.7cm}}
        \toprule
        \textbf{Model} & \multicolumn{2}{c}{\textbf{Dataset}} \\
        \midrule
        & \textbf{FinQA} & \textbf{ConvFinQA} \\
        \midrule
        \textit{\textbf{Finetuning Based}} \\ 
        \midrule
        GPT-2 & - & 59.12 \\
        T5 & - & 58.38 \\
        Retriever+NeRd & 48.57 & - \\
        Longformer & 21.90 & - \\
        FinQANet & 61.24 & 68.32 \\
        ELASTIC & 62.16 & - \\
        DyRRen & 63.30 & - \\
        ArgRecog & 64.86 & 73.94 \\
        TabT5 & 70.79 & - \\
        CellRetriever+UniLM & 68.00 & - \\
        ENCORE & 69.40 & 76.00\\

        \midrule
        \midrule
        \textit{\textbf{Prompting Based}} \\
        \midrule
        BloombergGPT & - & 43.41 \\
        GPT-3 Direct & 14.40 & 29.10 \\
        GPT-3 CoT & 26.10 & 37.40 \\ 
        GPT-3.5-turbo & 48.56 & 59.86 \\
        Codex Direct & 25.60 & 40.00 \\
        Codex CoT & 40.40 & 45.60 \\
        Codex CoT-SC & 44.40 & 47.90 \\
        PoT-Codex & 64.50 & 64.60 \\
        PoT-SC-Codex & 68.10 & 67.30 \\
        PoT-GPT-4 & 69.38 & 74.77 \\
        CBR-Ren & 67.81 & 72.61 \\
        \midrule
        \midrule
        \textit{\textbf{State of the Art}} \\
        \midrule
        APOLLO & \underline{71.07} & \underline{78.76} \\
        \textbf{\modelname} & \textbf{75.32} & \textbf{81.95} \\
        \midrule
        \midrule
        \textit{\textbf{Human Performance}} \\
        \midrule
        General Crowd & 50.68 & - \\
        Human Expert & 91.16 & - \\
        \bottomrule
      \end{tabular}
   }
\caption{Performance evaluation based on execution accuracy for FinQA and ConvFinQA datasets. The best performance is highlighted in bold, and the strongest baseline result is underlined. The baseline results are reproduced from APOLLO \cite{sun2024apollo} and PoT \cite{chen2023program}.}

   \label{table:result_table}
\end{table}

\fi

\begin{table}[!th]
   \scalebox{0.73}{
     \begin{tabular}{p{3.61 cm}|>{\centering\arraybackslash}p{1.6cm}|>{\centering\arraybackslash}p{1.6cm}|>{\centering\arraybackslash}p{1.6cm}}
        \toprule
        \textbf{Model} & \textbf{FinQA (dev)} & \textbf{FinQA (test)} & \textbf{ConvFinQA (dev)} \\
        \midrule
        \textit{\textbf{Finetuning Based}} \\ 
        \midrule
        GPT-2 & - & - & 59.12 \\
        T5 & - & - & 58.38 \\
        Retriever+NeRd & 47.53 & 48.57 & - \\
        Longformer & 23.83 & 21.90 & - \\
        FinQANet & 61.22 & 61.24 & 68.32 \\
        ELASTIC & 65.00 & 62.16 & - \\
        DyRRen & 66.82 & 63.30 & - \\
        ArgRecog & 67.50 & 64.86 & 73.94 \\
        TabT5 & - & 70.79 & - \\
        CellRetriever+UniLM & - & 68.00 & - \\
        ENCORE & 71.6 & 69.40 & 76.00\\

        \midrule
        \textit{\textbf{Prompting Based}} \\
        \midrule
        BloombergGPT & - & - & 43.41 \\
        GPT-3 Direct & - & 14.40 & 29.10 \\
        GPT-3 CoT & - & 26.10 & 37.40 \\ 
        GPT-3.5-turbo & - & 48.56 & 59.86 \\
        Codex Direct & - & 25.60 & 40.00 \\
        Codex CoT & - & 40.40 & 45.60 \\
        Codex CoT-SC & - & 44.40 & 47.90 \\
        PoT-Codex & - & 64.50 & 64.60 \\
        PoT-SC-Codex & - & 68.10 & 67.30 \\
        PoT-GPT-4 & 71.05 & 69.38 & 74.77 \\
        CBR-Ren & 68.40 & 67.81 & 72.61 \\
        
        \midrule
        \textit{\textbf{State of the Art}} \\
        \midrule
        APOLLO & \underline{72.91} & \underline{71.07} & \underline{78.76} \\
        \textbf{\modelname} & \textbf{77.13} & \textbf{75.32} & \textbf{81.95} \\
        
        \midrule
        \textit{\textbf{Human Performance}} \\
        \midrule
        General Crowd & - & 50.68 & - \\
        Human Expert & - & 91.16 & - \\
        \bottomrule
      \end{tabular}
   }
\caption{Performance evaluation based on execution accuracy for FinQA and ConvFinQA datasets. The best performance is highlighted in \textbf{bold}, and the strongest baseline result is \underline{underlined}. The baseline results are reproduced from APOLLO \cite{sun2024apollo} and PoT \cite{chen2023program}.}
\label{table:result_table}
\end{table}

\section{Results and Analysis}\label{sec:result_analysis}
We report the performance of our proposed model \modelname~ in Table~\ref{table:result_table}, including results on both the dev and test sets of the FinQA, and the dev set of the ConvFinQA. We observe that our approach outperforms the current best model, APOLLO \cite{sun2024apollo}, with realtive gains of \textbf{5.98\%} and \textbf{5.78\%} on the FinQA test and dev sets, respectively, and a \textbf{4.05\%} improvement on the ConvFinQA dev set. When compared with the best baseline under prompting based methods (PoT-GPT-4), our model shows \textbf{8.56\%} improvement on FinQA (test) and \textbf{9.60\%} improvement on ConvFinQA (dev). 

\noindent Codex variants generally performed lower. Among fine-tuning based methods, ENCORE \cite{wang-etal-2024-enhancing-numerical} performs the best on FinQA. Interestingly the execution accuracy for ConvFinQA is better than FinQA.
Providing questions in a conversational format helps the LLM follow a step-by-step approach, enhancing execution accuracy. 


\noindent\textbf{Comparison with ENCORE:} ENCORE \cite{wang-etal-2024-enhancing-numerical} reports SOTA on FinQA and ConvFinQA but underestimates APOLLO’s performance. Our model, \modelname, surpasses both APOLLO and ENCORE, achieving a new SOTA.

\noindent{\textbf{Analysis of Retriever Modules:}}
\label{sec:retriever_analysis}
In addition to FLAN-T5, we experimented with Mistral and APOLLO retrievers. As seen in Table~\ref{table:retriever_comparison} FLAN-T5 outperforms Mistral in cosine similarity despite having fewer parameters. We believe this is because Mistral generates additional lines and uses different wording, supported by the higher word count in its predictions (Table \ref{table:retriever_comparison}). Even the APOLLO retriever shows lower cosine similarity with ground truth facts, providing more irrelevant information, indicated by its higher word count. Our model (FLAN-T5) provides targeted retrieval with fewer irrelevant facts, indicated by its average word count that closely aligns with ground truth facts in FinQA and ConvFinQA (60.16 and 61.54, respectively). Our proposed retriever reduces confusion, helping the generator focus on relevant facts and improving execution accuracy (Table \ref{table:retriever performances}).

\begin{table}[!th]

  \scalebox{0.67}{
\begin{tabular}{c|cc|cc|p{1cm}}
\toprule
\textbf{Retriever} & \multicolumn{2}{c|}{\textbf{Cosine Similarity}} & \multicolumn{2}{c|}{\textbf{Average Words}} & \textbf{\#TP (M)} \\
\cmidrule(lr){2-3} \cmidrule(lr){4-5}
                   & \textbf{FinQA} & \textbf{ConvFinQA} & \textbf{FinQA} & \textbf{ConvFinQA} & \\
\midrule
Mistral            & 0.81           & 0.85               & 105.43         & 103.21             & 3.4 \\
FLAN-T5            & \textbf{0.91}  & \textbf{0.89}      & 88.60          & 74.32              & \textbf{0.59} \\
APOLLO             & 0.83           & 0.86               & 107.08         & 105.23             & 355 \\
\bottomrule
\end{tabular}
 }
 \caption{Analysis of retrievers. Cosine similarity and average number of words of the retrieved facts are shown. The last column TP denotes the approximate number of trainable parameters (in millions), highlighting the parameter efficiency of FLAN-T5.}
\label{table:retriever_comparison}
\end{table}

\subsection{Parameter Efficiency} 

The exceptional parameter efficiency of our FLAN-T5 retriever is demonstrated in the last column of Table~\ref{table:retriever_comparison}. By employing LoRA \cite{hu2021lora}, our retriever requires only 0.59 million trainable parameters, far fewer than the state-of-the-art retriever APOLLO, which requires 355 million parameters. This represents a compression ratio of nearly 600:1 while maintaining superior performance.

\section{Ablation Study}
We now try out various ablations over our proposed model to understand the significance of different modules, effect of in-context example selection etc. All ablation experiments are conducted on the FinQA test set and the ConvFinQA dev set.

\begin{table}[!ht]

 \small
      \resizebox{\linewidth}{!}{
      \begin{tabular}{l|c|c}
        \toprule
        Retriever & \multicolumn{2}{c}{Exe Accuracy} \\
        \cmidrule(lr){2-3}
                       & FinQA & ConvFinQA \\
        \midrule
        FLAN-T5        & \textbf{75.32} & \textbf{81.95} \\
        APOLLO        & 71.07 & 78.76 \\
        Mistral        & 72.88 & {80.81} \\
        APOLLO with ours strategy  & 72.97 & 80.67 \\

        \bottomrule
      \end{tabular} 
        }
          \caption{Results for our ablation studies of model performance with different retriever modules}
   \label{table:retriever performances}

\end{table}

\subsection{Retriever Modules' Performance}
\label{subsec:retriever_performance}
.
We compare the retriever performance, including final accuracy for FLAN-T5, APOLLO, and Mistral Table~\ref{table:retriever performances}. Despite being a larger LLM, Mistral underperforms compared to FLAN-T5 due to excessive irrelevant information, which confuses GPT-4 (see Section~\ref{sec:retriever_analysis}). Notably, all retrievers with our target computation surpass APOLLO, demonstrating the effectiveness of our approach.

In Table~\ref{table:retriever performances}, the last row shows APOLLO with our target computation surpassing its original version (second row) but falls short \modelname, validating the effectiveness of our framework.  

\subsection{Effect of In-context Examples Selection}
We experiment with different selection strategies of in-context examples and report the performance in Table~\ref{table:ablation_study_icl_examples}. 
We start with static examples, followed by vanilla PromptPG, which randomly selects examples instead of using clustered samples. Next, we experiment with a hybrid approach, combining two most similar examples with two selected by the PromptPG policy in a four-shot setting. Finally, inspired by \cite{lu2023dynamic}, our dynamic selection leverages clustering for strategic initialization, achieving the best performance across both datasets. Our experiment with reversing the order of examples shows a performance drop compared to our final model (Table~\ref{table:ablation_study_icl_examples}), emphasizing the importance of placing the highest-scoring example last for the most relevant context near the question.

\begin{table}[ht]
\small
       \resizebox{\linewidth}{!}{
      \begin{tabular}{l|cc}
        \toprule
        Model & \multicolumn{2}{c} {} \\
        & FinQA & ConvFinQA \\
        \midrule

        \modelname & \textbf{75.32} & \textbf{81.95} \\
        \quad w/ static in-context examples & 71.84 & 78.62 \\
        \quad w/ reverse ordering of in-context examples & 74.10 & 80.76 \\
        \quad w/ random selection & 72.53 & 79.57\\
        \quad w/ hybrid selection & 72.97 & 80.29 \\

        \bottomrule
      \end{tabular} 
        }
   \caption{Performance comparison of different in-context example selection strategies}
   \label{table:ablation_study_icl_examples}
\end{table}

\begin{figure}[!thb]
    \centering
     \includegraphics[width=1\linewidth]{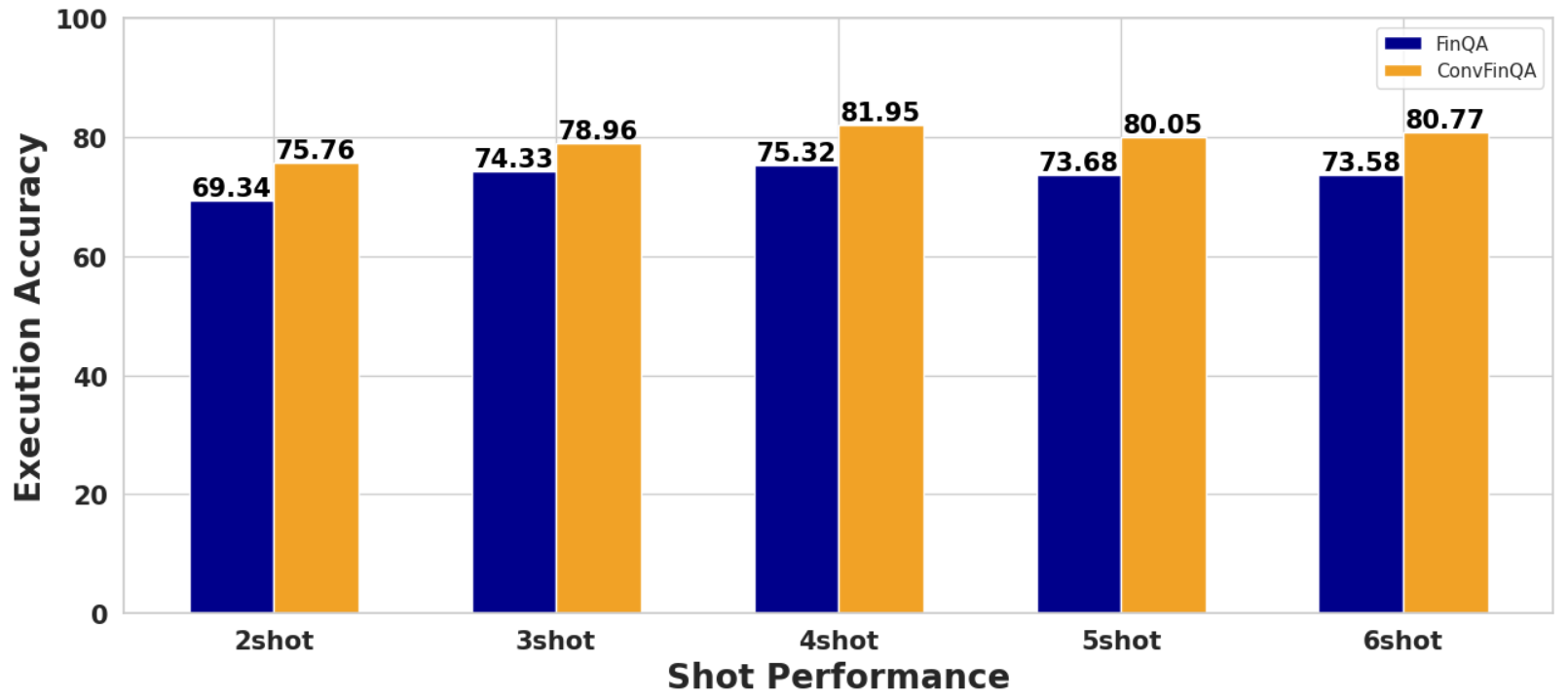}
    \caption{Few shot exemplar analysis of FinQA and ConvFinQA}
    \label{fig:few_shot_performance}
\end{figure}

\paragraph{Sensitivity to Exemplars:} To evaluate the sensitivity of our proposed model to different exemplars, we conducted a comprehensive few-shot analysis. For k-shot learning, we sampled k = (2, 3, 4, 5, 6) exemplars from the trained policy network. As shown in Figure \ref{fig:few_shot_performance}, we observe a significant drop in performance with 2 exemplars, followed by a relatively stable trend that peaks with 4 exemplars.

\begin{table}[ht]

\begin{tabular}{l|c|c}
        \toprule
         Question Types & \multicolumn{2}{c}{Dataset} \\
        \bottomrule
        & \multicolumn{1}{c|}{FinQA} & \multicolumn{1}{c}{ConvFinQA} \\
        \cmidrule{2-3}

Table-only questions & 81.02  & 88.19 \\
Text-only questions & 68.26  & 73.73 \\
Table-Text questions & 63.29  & 72.00 \\
Numeric questions & 75.24  & 82.17 \\
Boolean questions & 100.00  & 75.00 \\

\midrule

Program Steps \\ \hline
\bottomrule
 
        1 step programs & 80.58  & 84.17 \\
2 step programs & 72.61 & 78.79\\
>2 step programs & 52.38 & 72.73\\

\midrule
Programs with constants & 54.76 & 76.75\\
\bottomrule

\end{tabular}

\caption{Performance comparison on various Question Types and Program Steps}
\label{table:Question categorization}

\end{table}

\subsection{Evaluation with Other LLMs}
\label{subsec:other_llm_eval}
To evaluate the generalizability of the proposed \modelname~ pipeline, we conduct experiments using different LLMs for target computation module. In addition to GPT-4, we test with Gemini-2.0 \cite{gemini} and GPT-3.5-turbo \cite{openai2023gpt35} on the FinQA dataset. Even with the alternative LLM's, our approach surpasses the current state-of-the-art model APOLLO on the FinQA dataset.

\begin{table}[h!]
\setlength{\tabcolsep}{5pt} 
\centering
\resizebox{\linewidth}{!}{
\begin{tabular}{@{}l c@{}}
\toprule
Ours (LLM Engine) & Exec. Acc. (\%) \\
\midrule
\modelname~(GPT-4)         & \textbf{75.32} \\
\modelname~(Gemini-2.0)    & 74.89 \\
\modelname~(GPT-3.5-turbo) & 73.06 \\
\bottomrule
\end{tabular}
}
\caption{Evaluation of \modelname~with different LLMs on the FinQA (test) dataset.}
\label{tab:llm_comparison}
\end{table}

\subsection{Question Categorization vs Performance}
We evaluated model performance across five question types: table-only, text-only, table + text, boolean, and numeric (Table~\ref{table:Question categorization}). The model excels in table-only questions but struggles with table + text. For FinQA, it correctly answers all 20 boolean questions. While our framework lacks program steps, we analyze performance using ground truth steps. The model performs best on single-step programs, with multi-step and constant-based questions posing greater challenges.


\if{0}
\begin{figure}[ht]
        \centering
        \includegraphics[width=0.9\linewidth]{Images/IP_2009_45_1 (2).png}
        

        \caption{Logical error in sign conventions: The model calculates the signed difference instead of the absolute variation in lease obligations for 2011-2012.}
        \label{fig:IP-2009-page-45}
\end{figure}

\fi

\subsection{Error Analysis}  
\label{apn:error_analysis}
\begin{figure}[ht]
    \centering
        \includegraphics[width=\linewidth]{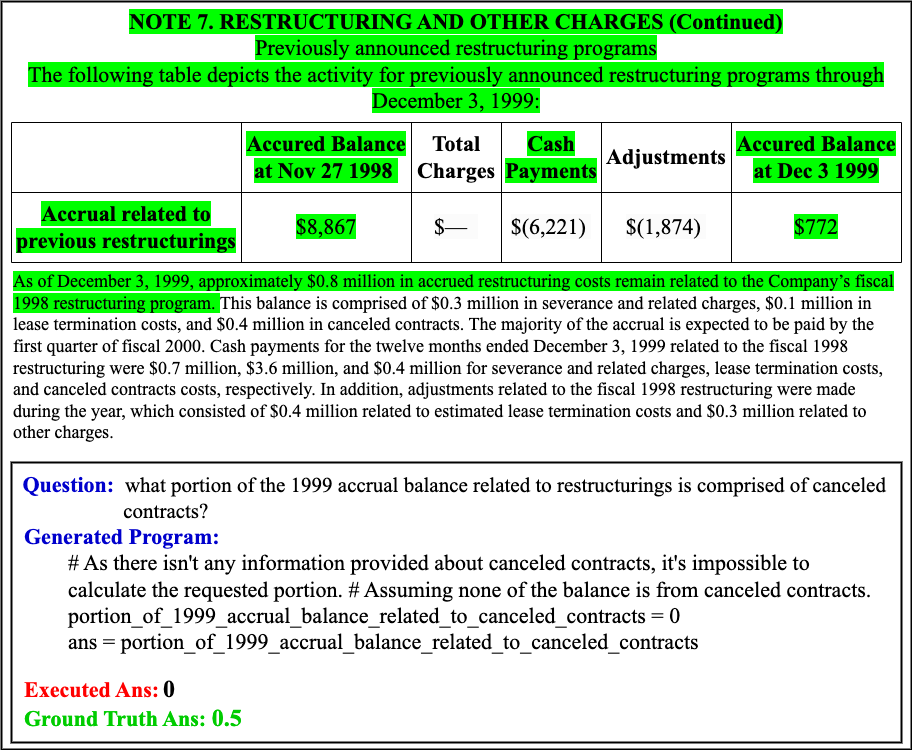}

        \caption{Fact Retrieval Errors: The Retriever module failed to retrieve the relevant facts necessary for the question.}
        \label{fig:ADBE-1999-page-64}
\end{figure}

We conducted a comprehensive error analysis by comparing the model's predictions against the ground truth to identify specific shortcomings. Errors were classified into three categories:

    \noindent \textbf{Fact Retrieval Errors}: Cases in which the model did not retrieve the correct facts required for the computation as shown in Figure ~\ref{fig:ADBE-1999-page-64}.

\begin{figure}[ht] 
    \centering
    \includegraphics[width=\linewidth]{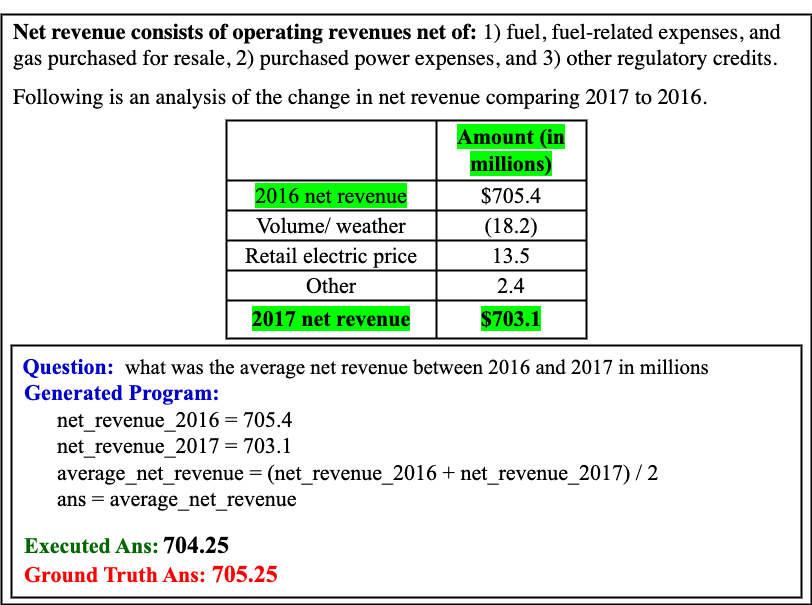}
    \caption{Wrong Ground Truth: Though the retrieved relevant facts, generated program, and executed answer are correct, there is a mistake in the ground truth answer. }
        \label{fig:ETR-2017-page-372}
\end{figure}

    \noindent \textbf{Ground Truth or Question Issues}: Instances where the ground truth was incorrect (refer to Figure~\ref{fig:ETR-2017-page-372}), incomplete, or ambiguous.  
 
  \noindent \textbf{Logical Errors}: Logical errors, the most common issue, stemmed from flawed reasoning or computation despite retrieving correct facts. For instance, the model miscalculated numerical differences by ignoring ground truth formats (Appendix \ref{apn:logical_error_difference}, Figure~\ref{fig:ECL-2017-page-94}) which represented the result as a percentage. or domain conventions (Figure~\ref{fig:IP-2009-page-45}). The program generated by the model, detailed in Figure~\ref{fig:IP-2009-page-45}, calculated the signed difference between the two values instead of the absolute variation.
  
A review of 100 FinQA errors found 9\% retrieval errors, 7\% ground truth issues, and the rest logical errors. Improving program generation, dataset annotations, and leveraging domain-specific knowledge can address these challenges.


\begin{figure}[ht]
        \centering
        \includegraphics[width=\linewidth]{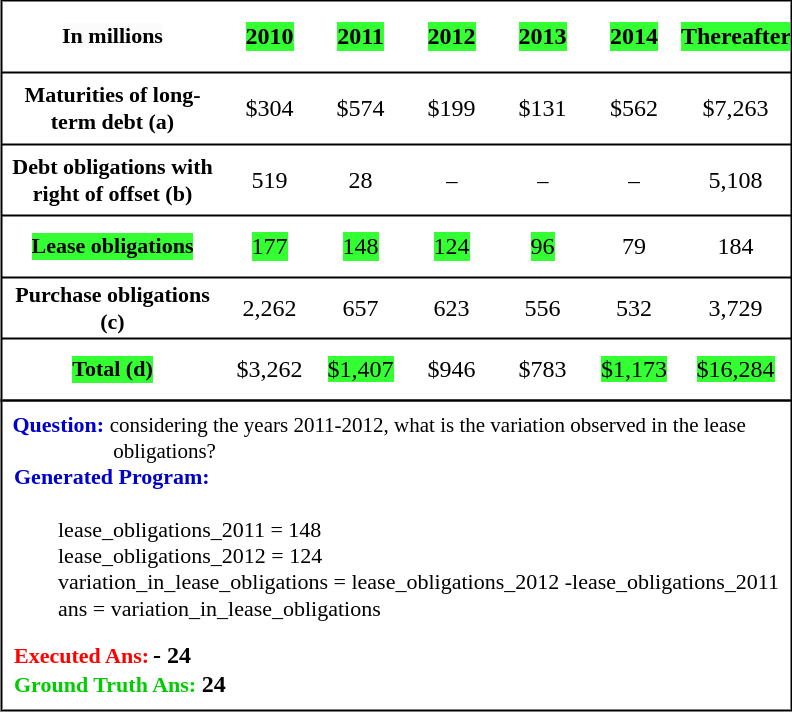}
        

        \caption{Logical error in sign conventions: The model calculates the signed difference instead of the absolute variation in lease obligations for 2011-2012.}
        \label{fig:IP-2009-page-45}
\end{figure}

\subsection{Comparative Analysis}
\label{apn:comp_analysis}


We conducted a manual analysis to compare the outputs of our target computation module, which leverages GPT-4 with program-of-thought-based prompting, under two conditions: when provided with all facts and data simultaneously, and when provided with facts retrieved by the APOLLO retriever. The analysis revealed that GPT-4 often makes grounding based errors when processing all facts at once. As shown in Appendix~\ref{apn:comp_analysis}, Table~\ref{tbl:finqa_comparison}, \modelname\ achieves better performance when provided with targeted facts rather than the full set of facts. This behavior is likely due to GPT-4's difficulty in extracting relevant facts effectively when faced with an excessive amount of information.
Furthermore, when integrating APOLLO with our target computation module, we observed cases where the facts retrieved by APOLLO were incorrect or incomplete (as illustrated in the first row of Table~\ref{tbl:finqa_comparison} in Appendix  \ref{apn:comp_analysis}). 
These inaccuracies in the retrieved facts ultimately lead to incorrect results and lower execution accuracy.



\if{0}
\subsection{Analysis of clustering with different embeddings}
We cluster the training pool of questions to ensure a diverse set of candidate in-context examples. Specifically, we experiment with Sentence-BERT and TF-IDF embeddings to generate representations of the questions, which are then clustered to select representative samples. Additionally, we explore a hybrid strategy that concatenates both embeddings. The execution accuracy of the \modelname~ pipeline using each of these clustering methods is reported in Table~\ref{tab:clustering-ablation}. We achieve best execution accuracy and silhouette score with the Sentence-BERT based representation.

\begin{table}[h]
\centering
\begin{tabular}{lc}
\toprule
\textbf{Clustering Strategy} & \textbf{Execution Accuracy (\%)} \\
\midrule
Sentence-BERT & \textbf{75.32} \\
TF-IDF        & 73.98 \\
Combination   & 74.34 \\
\bottomrule
\end{tabular}
\caption{Execution accuracy on FinQA using different clustering strategies for candidate in-context example selection.}
\label{tab:clustering-ablation}
\end{table}
\fi

%% file: Sections/Conclusion.tex
\section{Conclusion}
In this work, we introduced a novel two-step framework to enhance LLMs' ability to perform numerical reasoning in the financial domain. Our instruction tuned retriever,  accurately extracts relevant information from unstructured data sources. Our dynamic in-context example selection, guided by trained policy and clustering techniques, significantly improves the reasoning process. Future research will explore integrating external knowledge sources as well as to expand this framework to other domains needing complex numerical analysis. 

%% file: Sections/Limitations.tex
\section{Limitations}
In this work, we prioritized understanding the core reasoning capabilities of our model without incorporating external financial knowledge or human feedback. We did not experiment with models like GPT-4o \cite{hurst2024gpt}, which could be explored in future work. We did not report results on the ConvFinQA test set as the test set requires program submissions in a specific format, incompatible with our python based outputs. The interpretability of LLM reasoning processes and the potential propagation of errors within complex reasoning tasks requires further investigation. 



%% file: Sections/Appendix.tex
\section{Appendix}

\begin{table}[H]
     \centering
      \small
      \resizebox{\textwidth}{!}{

    \begin{tabular}{p{3cm} | p{4.5cm} | p{4.6cm} | p{4cm}}
        \toprule
        \textbf{Question} & \textbf{\modelname~Output} & \textbf{GPT-4 Output with all Facts} & \textbf{GPT4 + APOLLO Retriever} \\
        \midrule
        What is the borrowing under the term loan facility as a percentage of the total contractual maturities of long-term debt obligations due subsequent to December 31, 2016? & 
        borrowing\_under\_term\_loan = 3500\newline total\_contractual\_maturities = 150 + 175 + 2756\newline ans=borrowing\_under\_term\_loan/
        total\_contractual\_maturities\newline ans = 3500/3081 \textcolor{blue}{(Correct)} & 
        borrowing\_under\_term\_loan = 3500\newline total\_contractual\_maturities = 2014 + 150 + 175 + 2756\newline ans=borrowing\_under\_term\_loan/
        total\_contractual\_maturities\newline ans = 3500/5095 \textcolor{red}{(Wrong)\newline Target computation model given all facts but retrieved incorrect ones} & 
        borrowing\_under\_term\_loan = 3500\newline total\_contractual\_maturities = 2014 + 175 + 2756\newline ans=borrowing\_under\_term\_loan/
        total\_contractual\_maturities\newline ans = 3500/4945 \textcolor{red}{(Wrong)\newline Got wrong retrieved facts from APOLLO retriever} \\
        \midrule
        What is the total return if 100000 are invested in applied materials in 2008 and sold in 2011? & 
        investment\_2008 = 100000\newline value\_2008 = 100\newline value\_2011 = 118.21\newline ans = ((value\_2011/value\_2008) * investment\_2008) - investment\_2008\newline ans = 18210 \textcolor{blue}{(Correct)} & 
        investment\_2008 = 100000\newline value\_2008 = 100\newline value\_2011 = 118.21\newline ans = (value\_2011/value\_2008 * investment\_2008)\newline ans = 118210 \textcolor{red}{(Wrong)\newline Target computation model retrieved correct relevant facts but Incorrect logic} & 
        investment\_2008 = 100000\newline value\_2008 = 100\newline value\_2011 = 118.21\newline ans = (value\_2008/value\_2011 * investment\_2008)\newline ans = 0.11821 \textcolor{red}{(Wrong)\newline Got correct facts from APOLLO retriever but Incorrect logic} \\
        \bottomrule
    \end{tabular}
     }
        \caption{
Comparison of code outputs generated by \modelname, GPT-4, and APOLLO for a few sample questions, along with their respective final answers.}  


    \label{tbl:finqa_comparison}
\end{table}

\subsection{PoT results with GPT-4}
\label{apn:pot}
We re ran the POT paradigm code with the GPT-4 backbone. We achieved an execution accuracy of 69.38\% compared to the 74\% reported in the original work. The original PoT repository does not report GPT-4 responses on the FinQA test set, although it provides responses in json for other PoT variants. Even the current state-of-the-art APOLLO model does not report 74\% execution accuracy of PoT.

\subsection{Comparative Analysis}
\label{apn:comp_analysis}
 Our manual analysis (Table \ref{tbl:finqa_comparison}) highlights that GPT-4 performs better when guided by targeted facts retrieved by APOLLO, rather than processing all facts simultaneously. However, the effectiveness of this approach depends on the accuracy of APOLLO’s retrieval, as incorrect or incomplete facts can lead to reduced execution accuracy.

\subsection{Logical Error in Handling Differences}
\label{apn:logical_error_difference}
\begin{figure}[t]
        \includegraphics[width=0.9\linewidth]{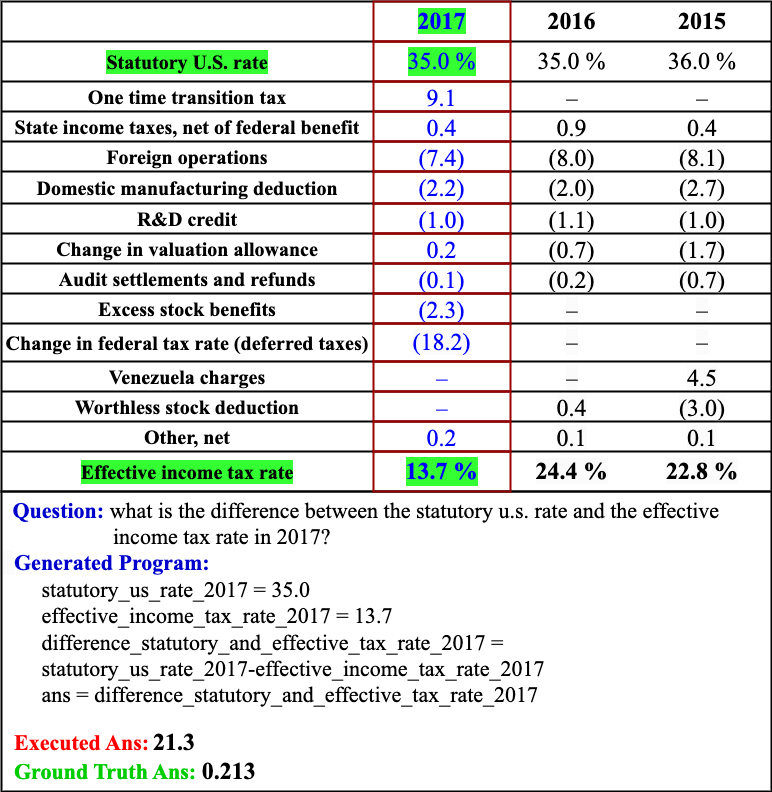}

        \caption{
Logical error in handling differences: The model computes the absolute numerical difference instead of representing the result as a percentage.}
        \label{fig:ECL-2017-page-94}


\end{figure}

In certain cases, the model exhibits a logical error when handling numerical differences. Specifically, instead of computing and representing the result as a percentage (\ref{fig:ECL-2017-page-94}).

